\documentclass[runningheads]{llncs}
\usepackage{graphicx}
\usepackage{xspace}
\usepackage{amsmath,amssymb,amsfonts}
\usepackage{algorithmic}
\usepackage{textcomp}
\usepackage{xcolor}
\usepackage[T1]{fontenc}
\usepackage{listings}
\usepackage{subfig}
\lstdefinestyle{mystyle}{frame=single,showstringspaces=false}    
\begin{document}
%

%%%%%%
% PI_CAST: CLIENT
% PI_UNITY: NODE1
% PI_SECONDARY: NODE2

% NETBOOK: CONTINUOUS ATTACK
% PARAMS

%%%%%%

\title{Search@Home: A Commercial Off-the-Shelf Environment for Investigating Optimization Problems}

\newcommand{\ToolName}{\textit{Search@Home}\xspace}
\newcommand{\etal}{\emph{et al.}}
\titlerunning{\textit{Search@Home}}%~\mbox{~}}
% If the paper title is too long for the running head, you can set
% an abbreviated paper title here
%
%\author{Authors removed for anonymous submission}
\author{Erik M. Fredericks\inst{1} \and Jared M. Moore\inst{2}}%\orcidID{0000-0003-4287-3339}} %\and
%Second Author\inst{2,3}\orcidID{1111-2222-3333-4444} \and
%Third Author\inst{3}\orcidID{2222--3333-4444-5555}}
%

%\authorrunning{Authors anonymized}
\authorrunning{E. Fredericks and J. Moore}

% First names are abbreviated in the running head.
% If there are more than two authors, 'et al.' is used.
%

%\institute{Institutions removed for anonymous submission}

\institute{Dept. of Computer Science \& Engineering, Oakland University\\Rochester MI, 48309, USA, 
\email{fredericks@oakland.edu}
\and
School of Computing and Information Systems, Grand Valley State University\\ Allendale MI, 49401, USA, 
\email{moorejar@gvsu.edu}}

%\and
%Springer Heidelberg, Tiergartenstr. 17, 69121 Heidelberg, Germany
%\\\email{EMAIL TBD}}\\
%\url{http://efredericks.net}}% \and
%ABC Institute, Rupert-Karls-University Heidelberg, %Heidelberg, Germany\\
%\email{\{abc,lncs\}@uni-heidelberg.de}}
%
\maketitle              % typeset the header of the contribution
%

% 192.168.1.101 - raspberrypi01 - haproxy
% 192.168.1.102 - raspberrypi02 - haproxy
% 192.168.1.105 - pi4 - haproxy master
% 192.168.1.110 - pinode1 - dispy?
% 192.168.1.111 - pinode2 - dispy?

\begin{abstract}
%Search-heuristics, particularly those that are evaluation-driven (such as evolutionary computation), are often performed in simulation environments that can execute quickly and enable parallel investigations of parameters at the cost of experiencing real-world situations.  However, search heuristics have been also proven to be successful in constrained environments with limited searching capacity even when the solution space is broad.  This paper introduces \ToolName: an environment that comprises heterogeneous, commercial off-the-shelf devices to enable rapid prototyping of real-world optimization problems.  Given its relative cost-effectiveness, \ToolName can also serve as a rich environment for introducing STEM students to the search-based domain.  We demonstrate its effectiveness with three empirical investigations at different levels of complexity.
%With recent advances in both cloud and microcomputer technologies, search 

%Search heuristics, particularly those that are evaluation-driven (e.g., evolutionary computation), are often performed in simulation to enable exploration of large solution spaces at the detriment of real-world conditions.  

Search heuristics, particularly those that are evaluation-driven (e.g., evolutionary computation), are often performed in simulation, enabling exploration of large solution spaces.  Yet simulation may not truly replicate real-world conditions. However, search heuristics have been proven to be successful when executed in real-world constrained environments that limit searching ability even with broad solution spaces.  Moreover, searching \textit{in situ} provides the added benefit of exposing the search heuristic to the exact conditions and uncertainties that the deployed application will face.  
Software engineering problems can benefit from \textit{in situ} search via instantiation and analysis in real-world environments.
This paper introduces \ToolName, an environment comprising heterogeneous commercial off-the-shelf devices enabling rapid prototyping of optimization strategies for real-world problems.

%We demonstrate its effectiveness with two empirical investigations at different levels of complexity: a string optimization problem and an open-source website load balancer that has been modeled as a software engineering optimization problem.

\end{abstract}
\section{Introduction}
%P1
%Local computing capabilities are increasing with the recent rise in inexpensive commercial off-the-shelf (COTS) microcomputers (e.g., Raspberry Pi, Arduino, BeagleBone, etc.\footnote{See \url{https://www.raspberrypi.org/}, \url{https://www.arduino.cc/}, and \url{https://beagleboard.org}, respectively.})  However, such devices are generally constrained in terms of available processing power and hard drive space, limiting their ability to carry out complicated tasks quickly.  A Raspberry Pi 3B, for instance, has a four-core ARM processor at 1.2GHz, with 1GB of available RAM~\cite{barnes.magpi.2016}.\footnote{The newest model, the Raspberry Pi 4B, has 8GB RAM available.}  While this seems reasonable in terms of processing power, there exist limitations in data speed transfers on the board itself, leading to bottlenecks that can impact compute jobs~\cite{barnes.magpi.2016}.  There are benefits in that a microcomputer will generally consume far less power than a typical device (e.g., server blade) that would be deployed for intensive computing tasks.  However, a microcomputing environment is attractive in terms of rapidly deploying and testing a search heuristic in real-world conditions, rather than relying on simulation. 
Commercial off-the-shelf (COTS) microcomputers (e.g., Raspberry Pi, Arduino, BeagleBone, etc.\footnote{See \url{https://www.raspberrypi.org/}, \url{https://www.arduino.cc/}, and \url{https://beagleboard.org}, respectively.}) are democratizing access to many-core distributed computing environments.  Such devices are generally constrained in terms of available processing power and hard drive space, limiting their ability to individually carry out complicated tasks quickly. 
%Local computing capabilities are increasing with the recent rise in inexpensive commercial off-the-shelf (COTS) microcomputers (e.g., Raspberry Pi, Arduino, BeagleBone, etc.\footnote{See \url{https://www.raspberrypi.org/}, \url{https://www.arduino.cc/}, and \url{https://beagleboard.org}, respectively.})  However, such devices are generally constrained in terms of available processing power and hard drive space, limiting their ability to carry out complicated tasks quickly.  
%A Raspberry Pi 3B, for instance, has a four-core ARM processor at 1.2GHz, with 1GB of available RAM~\cite{barnes.magpi.2016}.\footnote{The newest model, the Raspberry Pi 4B, has 8GB RAM available.}  While this seems reasonable in terms of processing power, there exist limitations in data speed transfers on the board itself, leading to bottlenecks that can impact compute jobs~\cite{barnes.magpi.2016}.  
%%%%%For example, a server blade environment can simulate devices in a smart home environment, however microcomputers can be readily deployed \textit{in situ} for experimentation purposes.  
Moreover, such devices will generally consume far less power than a typical device used for simulation (e.g., server blade).  We posit that a microcomputing environment is therefore beneficial in terms of rapidly prototyping search heuristics,  in real-world conditions, yet may require additional time to complete complex computing tasks.  
Specifically, we highlight search-based software engineering (SBSE) as an attractive domain for real-world search.

%P2
Optimization performed in real-world situations allows the system to analyze relevant information from data specific to its operating context (i.e., combination of system and environmental parameters) rather than simulating such parameters~\cite{bertsimas.2006}.  Self-adaptation has been applied to cyber-physical systems enabling reconfiguration at run time in response to uncertainty, including those systems considered to be safety critical~\cite{muccini.2016}.  Online evolutionary optimization has been previously performed in fields such as robotics, where a (1+1) evolutionary strategy searches for optimal neural network configurations~\cite{bredeche.2010}.  Given the difficulties in performing speculative, evaluation-driven optimizations at run time, continuous optimization methods such as Markov chains and Bayesian methods have also been applied~\cite{calinescu.2010,shahriari.2015}.  Regardless of the method, an online optimization strategy must consider the implications of updating an executing system within its production environment.

%P3
This paper introduces and demonstrates \ToolName, a framework for quickly prototyping in-place search-based software engineering (SBSE) techniques using COTS hardware. \ToolName is intended to provide a low-cost testbed that can be deployed in a target environment to rapidly prove out online search heuristics (e.g., run-time requirements monitoring/optimization).  While a longer evaluation time is to be expected with low-power hardware, the benefits of performing search in a real-world environment far outweigh the speed gains of using a simulation that may be misconfigured or inaccurately describe the environment-to-be.  We next describe background and related work, demonstrate \ToolName on a proof-of-concept optimization problem, and outline future experiments.
\section{Real-World Systems}
\label{sec:bg}
%This section discusses relevant background information on COTS microcomputers and online optimization.%, and load balancing.
This section presents relevant information on microcomputers and how SBSE techniques can be effective within constrained environments.

%\subsection{Microcomputers}

\vspace{0.1in}

\noindent \textbf{Microcomputers}: Consumer-grade microcomputers have been widely propagated at inexpensive price points with the recent explosion of interest in Maker-related topics (e.g., hacking home electronics, 3D printing, etc.), many of which are targeted at STEM education and non-production projects.  Moving from simulation to reality in constrained environments requires addressing concerns in:  %Numerous manufacturers exist to support both STEM and industrial needs, however for the purposes of this paper we will limit our discussion to the Raspberry Pi with comparisons to the Arduino and BeagleBone platforms.%\footnote{Other popular platforms include the Nvidia Jetson Nano (\url{https://www.nvidia.com/en-us/autonomous-machines/embedded-systems/jetson-nano/}), Asus Tinker Board (\url{https://www.asus.com/Single-Board-Computer/Tinker-Board/}), and BBC micro: bit (\url{https://microbit.org/}).}

%\vspace{0.2in}

%\noindent \textbf{\textit{Real-world concerns}}: We will now highlight several problems that arise when moving from simulation to reality in low-capacity environments:

\begin{itemize}
    \item \textbf{\textit{Power}}: Beyond power optimization, microcomputers often supplied by sub-standard power cables, resulting in \textit{undervoltage} (i.e., slow/erratic behavior).
    %slow processing speed and erratic behaviors from \textit{undervoltage}. 
    %\item \textbf{\textit{Power}}: While power optimization can be a concern in and of itself, microcomputers often are sold with sub-standard power cables.  Another possibility is that ``another cable'' is substituted from a home kit with the same connectors, albeit with different specifications.  As such, a device may experience \textit{undervoltage}, resulting in slower processing speeds or erratic behaviors.
     \item \textbf{\textit{Temperature}}: Heat can negatively impact devices without active (i.e., fans) or passive (i.e., heat sinks) cooling,
     %Heat dissipation arises as such devices may not have any form of active (i.e., cooling fans) or passive (i.e., heat sinks) cooling while performing strenuous tasks, 
     leading to CPU throttling.% (e.g., slow down) until temperatures normalize.
     %leading to the microcomputer's processor operating at dangerous temperatures.  This issue often results in CPU throttling (e.g., slow down)  until the device cools to acceptable temperatures.
   
    %\item \textbf{\textit{Temperature}}: Another concern for microcomputers is heat dissipation.  Such devices are often set up without any form of either active (e.g., cooling fans) or passive (e.g., heatsinks) cooling devices, leading to a state in which the microcomputer's processor is operating dangerous temperatures.  This state often results in CPU throttling, or slowing down processing speed until the device cools to acceptable temperatures.
    \item \textbf{\textit{Memory}}: Microcomputers are constrained with the amount of available memory for handling computing tasks.  While the newest Raspberry Pi (4B) has up to 8GB of available RAM, older models have significantly less memory.  
    %Moreover, devices such as the Arduino Uno have 32KB of flash memory available,\footnote{See \url{https://www.arduino.cc/en/Tutorial/Memory}.} resulting in lower processing speeds. 
    Edge devices exist specifically for heavy-duty computing (e.g., Google Coral\footnote{See \url{https://www.coral.ai/}.}), however they are generally used for only specialized purposes.  
    %\item \textbf{\textit{Memory}}: Microcomputers are often constrained with the amount of available memory for handling computing tasks.  While the newest Raspberry Pi (4) has 4GB of available RAM, older models have significantly less.  Moreover, devices such as the Arduino Uno have 32KB of flash memory available,\footnote{See \url{https://www.arduino.cc/en/Tutorial/Memory}.} resulting in even less available processing capabilities.  While edge devices exist specifically for heavy-duty computing (e.g., Google Coral\footnote{See \url{https://www.coral.ai/}.}), such devices typically are not often found in COTS setups unless if there was a specific need.  
    \item \textbf{\textit{Disk space}}: Depending on the device, permanent storage (e.g., EEPROM, ROM, flash memory, etc.) is often at a premium in terms of availability. Devices such as the Arduino and BeagleBone rely on programmable memory space for long-term storage, whereas the Raspberry Pi uses a microSD card for its storage.
   %\item \textbf{\textit{Networking}}: \ToolName requires a network of devices, and as such, each device must have an accessible \textit{route} to all other devices that are participating.  One particular issue that can arise is that IP addresses are generally automatically assigned via DHCP, meaning that the IP address can change when its lease time expires.  Such an event can cause catastrophic failure if IP addresses are hard-coded into configuration files.
    \item \textbf{\textit{Timing constraints}}: While the devices used in this paper do not use real-time operating systems, timing constraints must be explicitly handled by the engineer, else faults can occur when software deadlines are violated.
    
\end{itemize}

%\subsection{Online Optimization}

\noindent \textbf{Online/hardware-based optimization}: One common theme across optimization algorithms is that there exists no ``free lunch'' as there are always limiting factors in the application or environment~\cite{wolpert1997no}.  
Limitations for online optimization will be readily-noticeable  in run time, memory overhead, etc.  In constrained systems these impacts are exceedingly noticeable given their lower operating capabilities.  Care must be taken when performing optimization in constrained systems.

\textit{In situ} optimization has been applied to wireless sensor network applications, where run-time reconfiguration and programming models enable optimization~\cite{taherkordi.2013}.  
%Fredericks~\etal{} combined evolutionary computation with continuous optimization and clustering methods to optimize support planning in self-adaptive systems~\cite{fredericks.2019.saso}. %\textit{In situ} optimization has also been applied to the software engineering domain. 
Li~\etal{} minimized power consumption of test suites in low power systems using an integrated linear programming approach~\cite{li.2014}.
Continuous optimization is a common feature in other domains as well.  Wang and Boyd applied an online optimization technique (a primal barrier method) to model-predictive control applications~\cite{wang.2009.cst}.  Mars and Hundt combined static and dynamic optimization strategies to direct online reconfiguration based on scenarios~\cite{mars.2009}.  Optimization has also been applied online in a data-driven capacity, where uncertainty models can inform  decision making~\cite{bertsimas.2006}.

Search heuristics have also been deployed \textit{in silica}. Genetic algorithms (GA) have been deployed to field-programmable gate arrays (FPGA) for hardware-based optimization~\cite{peker.2018},
%guo.2015,
enabling rapid prototyping of fast, low-power search.  Energy management is another concern for metaheuristics, specifically those involving smart power grids~\cite{lezama.2018}.  With additional hardware modules, \ToolName can be extended to use \textit{in silica} search and act instead as a controller, and moreover, monitor energy consumption with appropriate sensors.

\section{\textit{Search@Home} Overview}%Implementation}
\label{sec:approach}
%This section describes our implementation of \ToolName , an overview of the string search problem, and possibilities for STEM outreach.
%and the deployment of \ToolName in a load-balanced network application.

%\subsection{\textit{Search@Home} Overview}

For the purposes of this paper (and its initial implementation), \ToolName was implemented as an IoT environment comprising two Raspberry Pi 3B devices, a Raspberry Pi 4B (4GB model), a spare 2010-era netbook (acting as a point of entry to the network), and a wireless router (flashed with Tomato firmware) 
%Tomato custom firmware~\cite{tomato.firmware}) 
to provide a sandboxed network.  Devices can be interchangeable -- as long as the device can connect to the network it can participate in the environment.  For example, an Arduino Duo could be included as long as an appropriate communication channel to the rest of the network is established (i.e., a WiFi shield is installed).

%PICTURE OF ENV

We applied a string optimization problem demonstrating the feasibility of deploying a search algorithm to a heterogeneous collection of devices.  The algorithm was developed with Python 3.7 and each Raspberry Pi used Raspbian as its operating system.  Given that Raspbian is a full Linux-based operating system, libraries such as \texttt{DEAP} and \texttt{DisPy} can facilitate development of multiple search heuristics or distributed cluster computing, respectively.\footnote{See \url{https://deap.readthedocs.io/} and \url{http://dispy.sourceforge.net/}, respectively.} 
To allow for replication, we have made our parts list, source code, and results publicly available on GitHub.\footnote{See https://github.com/efredericks/SearchAtHome.}
%relevant configurations, 

%\ToolName was applied to two application domains to demonstrate its feasibility: string search and a load-balanced networking application.  The search algorithms were developed using Python 3.7 for both applications.  However, given that a Raspberry Pi runs a full Linux operating system, libraries such as \texttt{DEAP}\footnote{See \url{https://deap.readthedocs.io/}.}, as well as other languages (e.g., C++, Java) are feasible.  The string search algorithm was performed on a single Raspberry Pi device (both on the Pi 3B and Pi 4B for comparative purposes).  A distributed/parallel computing framework such as \texttt{dispy}\footnote{See \url{http://dispy.sourceforge.net/}.} is feasible as well for a more computation-heavy application.

%For a distributed application (i.e., our load balancer), the Raspberry Pi 4B was considered to be the \textit{master} node in charge of the search algorithm.  For the purposes of the \texttt{HAProxy} load balancer (see Section~\ref{sec:haproxy}), this node was also responsible for acting as the web server frontend.\footnote{These applications could be separated, however the Pi 4B was capable of running both applications without negative impact.}  

%}See the \texttt{ssbse20} branch: \url{https://github.com/efredericks/HomeLab}.}

%\subsection{String Search}
%\label{sec:string}

\vspace{0.1in}

\noindent \textbf{String search configuration}: 
%To demonstrate the feasibility of running search on a constrained device, we implemented a  string search to baseline the performance and reliability of a Raspberry Pi. 
The string search application leverages a standard GA to search for a given string, where this particular GA was released as a GitHub.\footnote{See \url{https://gist.github.com/josephmisiti/940cee03c97f031188ba7eac74d03a4f}.}
%~\cite{drake.ga.2016}.  
To increase the difficulty of this task, we specified a complex string (\texttt{OPT}) comprising \texttt{ASCII} characters sampled between indices $32$ and $64$ and a total length of $84$.\footnote{There exist $2.7 * 10^{126}$ possible combinations based on string length and characters.}  \texttt{OPT} is defined in Equation~\ref{eq:1}:

\vspace{-0.15in}

\begin{align}
\begin{split}\label{eq:1}
    OPT ={}& \textquotesingle 12th\mbox{~}Symposium\mbox{~}for\mbox{~}Search-Based\mbox{~}Software\\
         & Engineering\mbox{~}|\mbox{~}http://ssbse2020.di.uniba.it/ \textquotesingle
\end{split}
\end{align}

The fitness function defined for this study is defined as follows in Equation~\ref{eq:2}, where \texttt{ord} represents a character's Unicode integer value:

\vspace{-0.10in}

\begin{equation}
    ff_{string} = \sum_{i=0}^{len(OPT)} |ord(TARGET[i]) - ord(OPT[i])|
\label{eq:2}
\end{equation}

This particular GA uses single-point crossover,  single-point mutation, and a weighted-fitness selection function (i.e., fitness directly correlated to the probability that an individual is selected).  Mutation is automatically applied to the children generated by the crossover operation, where a random gene is modified.  The GA was configured to run for a maximum of $500,000$ generations, with a population size of $500$, a crossover rate of $0.5$, and a mutation rate of $1.0$.  The GA can converge early if the correct string is discovered.  We performed $25$ experimental replicates to ensure statistical significance, using the Wilcoxon-Mann-Whitney U-test with a significance level of $p$ $<$ $0.001$.  We compared the GA to random search as specified by Arcuri~\etal{}~\cite{arcuri.2011}.

%The evolutionary parameters are configured as specified in Table~\ref{table:simple-ga-params}:

%\begin{table}[ht]
%\centering
%\caption{String Search Genetic Algorithm Parameters.}
%\begin{tabular}{|l|c|}
%\hline
%\textbf{Operator} & \textbf{Value} \\ \hline
%Number of (maximum) generations & $500000$ \\ \hline
%Population size & $500$ \\ \hline
%Crossover rate & $0.5$ \\ \hline
%Mutation rate & $1.0$ \\ \hline
%\end{tabular}
%\label{table:simple-ga-params}
%\end{table}

%Note that, for this particular problem, the GA can converge early (i.e., before the number of generations is reached) if the optimal string value is discovered.  
%We performed $25$ experimental replicates (i.e., each replicate is a repeatably-seeded evaluation) to ensure statistical significance.  We also used the Wilcoxon-Mann-Whitney U-test with a significance level of $0.001$ (i.e., $p < 0.001$).  Furthermore, the GA was compared to random search to demonstrate its effectiveness, as specified by Arcuri~\etal{}~\cite{arcuri.2011}.

\begin{figure}[htb!]%
    \centering
    \subfloat[Number of generations.]{{\includegraphics[width=5.5cm]{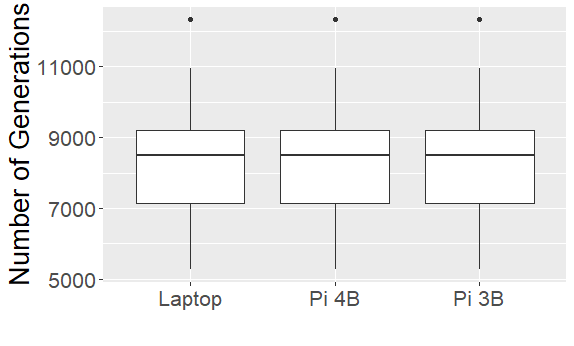}}} %\includegraphics[width=5cm]{img1} }}%
    \qquad
    \subfloat[Execution time.]{{\includegraphics[width=5.5cm]{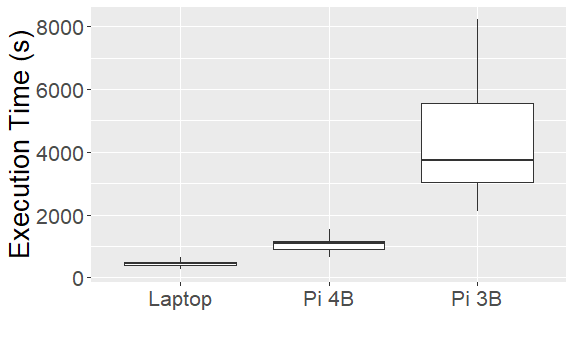}}} %\includegraphics[width=5cm]{img2} }}%
    
    \vspace{-0.10in}
    
    \caption{Comparison of string search results between laptop and Raspberry Pis.}%
    \label{fig:stringsearch}%
\end{figure}

For this feasibility study, we are interested in the amount of time required before convergence to the expected solution and the number of generations necessary to reach that value, as the number of generations is set very high to ensure convergence.  
%We compare these values to those performed on a current-generation laptop Intel Core i7 quad-core 2.8GHz 64-bit processor, 16GB RAM, 1TB hard drive space).  The microcomputers used for this experiment were a Raspberry Pi 3B with an ARM quad-core 1.2GHz 64-bit processor, 1GB of RAM and 16GB of hard drive space and a Raspberry Pi 4B with an ARM quad-core 1.5GHz 64-bit processor, 4GB of RAM, and 16GB of hard drive space.  
To focus on execution time, we intentionally did not introduce parallelism or distributed processing, however such a procedure can be used on a microcomputer (e.g., Python's \texttt{multiprocessing} package). %\footnote{Both Raspberry Pis have 4 processor cores and could leverage Python's \texttt{multiprocessing} package, for example.}
 Figure~\ref{fig:stringsearch}(a) compares the number of generations required for the algorithm to converge and Figure~\ref{fig:stringsearch}(b) compares the amount of time (seconds) required to reach convergence between a current-generation laptop\footnote{Intel Core i7 quad-core 2.8GHz 64-bit processor, 16GB RAM, 1TB hard drive space.} and the Raspberry Pis.  As can be seen from these figures, there exists no difference in the number of generations required to converge to the optimal solution.  However, a significant difference exists between the execution times required for convergence between each device ($p < 0.001$).  Moreover, each experimental replicate resulted in convergence.  These results are expected given the disparity in processing capability.  However, the Raspberry Pi was able to successfully execute optimization in all cases within a reasonable amount of time, proving feasibility of optimization on constrained devices.

\vspace{-0.25in}

\noindent \textbf{SBSE Implications}: We now highlight three future experiments for \textit{in situ} optimization research.   

%% Possible projects
\begin{enumerate}
    \item Distributed processing of search algorithms
    \item Power concerns resulting from search
    \item Implications of SBSE in production environments
\end{enumerate}

%While each item listed above can be a research project in and of itself, we feel that highlighting future topics as possible STEM coursework can be of benefit to the SSBSE community.  Item (1) can be an interesting study of technologies such as a Beowulf cluster, HPCC deployment, or \texttt{DisPy} environment\footnote{See \url{http://dispy.sourceforge.net/}.} in which IoT devices are setup as nodes in a cluster for executing local search, where nodes may be tasked with both fitness evaluation and distributed evolutionary operations.  Item (2) presents interesting implications in modeling power consumption concerns as a first-class citizen in the search process, possibly modeled as non-functional requirements, given that a COTS deployment often relies on home utility services.  Item (3) relates to the use of \ToolName \textit{in situ}, where continuously updating system configurations as a result of search impacts the overall behavior of the deployed system.  
Item (1) can be an interesting study of offloading SBSE tasks (e.g., requirements monitoring, fitness evaluation, etc.) to distributed nodes.  Item (2) demonstrates the implications of modeling power consumption as a first-class citizen in a software model (e.g., non-functional requirements).  Item (3) uses optimization \textit{in situ} to investigate the interaction of search, software artifacts, and expressed behaviors.
%While each item listed above can be a research project in and of itself, we feel that highlighting future topics as possible STEM coursework can be of benefit to the SSBSE community.  Item (1) can be an interesting study of technologies such as a Beowulf cluster, HPCC deployment, or \texttt{DisPy} environment\footnote{See \url{http://dispy.sourceforge.net/}.} in which IoT devices are setup as nodes in a cluster for executing local search, where nodes may be tasked with both fitness evaluation and distributed evolutionary operations.  Item (2) presents interesting implications in modeling power consumption concerns as a first-class citizen in the search process, possibly modeled as non-functional requirements, given that a COTS deployment often relies on home utility services.  Item (3) relates to the use of \ToolName \textit{in situ}, where continuously updating system configurations as a result of search impacts the overall behavior of the deployed system.  

%\section{Proof of Concept Results}
%\label{sec:expr}
%\input{sections/04-results}

%\section{Related Work}
%\label{sec:rel}
%\input{sections/05-related-work}

\section{Discussion}
\label{sec:discussion}
%In this paper, we have presented \ToolName, a design-time and run-time approach to explicitly model enumerated environmental scenarios, verify expected system operation in the context of those environmental scenarios, and detect unexpected environmental scenarios at run time.

%We demonstrated \ToolName on two sets of goal models that describe baby monitors of differing complexities. We showed that \ToolName is able to automatically detect environmental scenarios outside of the explicitly defined environmental scenarios in the environmental model in order to determine if the system had been verified at design time for that scenario.

%Future research directions include exploration how \ToolName can be extended to address temporal and RELAXed properties~\cite{whittle2009relax,whittle2010relax}, as well as the optimal distribution of limited sensors for multi-system analysis.

This paper presents \ToolName, an open-source framework for enabling \textit{in situ} SBSE research within constrained environments.  \ToolName uses inexpensive COTS hardware providing an environment in which students and researchers can quickly and effectively prototype and deploy applications that would benefit from using real-world data to support an online search procedure.  For this paper, we targeted evolutionary computation, however extension to other optimization domains (e.g., continuous optimization) is feasible as well.

We demonstrate the effectiveness of \ToolName on a string search exemplar to demonstrate basic search feasibility, where optimal solutions are discovered in a reasonable amount of time.  Future research directions for this project include incorporation of low-cost robotics environments (e.g., iRobot Roomba, Turtlebot, Lego Mindstorms, etc.), usage of a compute cluster (e.g., Beowulf cluster, high-performance compute cluster, etc.), and incorporation of cloud technologies offsetting the cost of heavy evaluations (e.g., Google Cloud Functions, Amazon Web Services Lambda functions, etc.).

\vspace{0.1in}

\noindent \textbf{Acknowledgements}. This work has been supported in part by grants from the National Science Foundation (CNS-1657061),  Oakland University, and Grand Valley State University.  The views and conclusions contained herein are solely those of the authors.

%\vspace{-0.15in}

%\section*{Acknowledgements}

%This work has been supported in part by grants from the National Science Foundation (CNS-1657061),  Oakland University, and Grand Valley State University.  The views and conclusions contained herein are solely those of the authors.% and should not be interpreted as necessarily representing the official policies or endorsements, either expressed or implied, of the National Science Foundation or Oakland University.

% ---- Bibliography ----
%
% BibTeX users should specify bibliography style 'splncs04'.
% References will then be sorted and formatted in the correct style.
%
% \bibliographystyle{splncs04}
% \bibliography{mybibliography}
%
\bibliographystyle{splncs04}
\bibliography{efredericks_master.bib}
\end{document}